# Semantic Linking –
# a Context-Based Approach to Interactivity in Hypermedia

Michael Engelhardt, Thomas C. Schmidt

Hochschulrechenzentrum
Fachhochschule für Technik und Wirtschaft Berlin
Treskowallee 8
D-10318 Berlin
engelh@fhtw-berlin.de
schmidt@fhtw-berlin.de

**Abstract:** The semantic Web initiates new, high level access schemes to online content and applications. One area of superior need for a redefined content exploration is given by on-line educational applications and their concepts of interactivity in the framework of open hypermedia systems.

In the present paper we discuss aspects and opportunities of gaining interactivity schemes from semantic notions of components. A transition from standard educational annotation to semantic statements of hyperlinks is discussed. Further on we introduce the concept of semantic link contexts as an approach to manage a coherent rhetoric of linking. A practical implementation is introduced, as well.

Our semantic hyperlink implementation is based on the more general Multimedia Information Repository MIR, an open hypermedia system supporting the standards XML, Corba and JNDI.

## 1   Introduction

The semantic web recently was created as an initiative to bring machine congestible structure to the bulk of Web information, which had been designed only for human reception. A vision was presented of robots and crawlers digesting online material on a level sufficient to be positioned as 'new' interfaces acting between content and human, providing superior navigational knowledge.

Navigational intelligence, in contrast, has been a subject of lengthy investigations in the community of open hypermedia systems, adjusting the focus on options and requirements of applications. The field of hypermedia has been freshly inspired by manifold open and distance learning activities and meta description standards, as well.

Bringing those fields together a view opens up on human susceptible applications grounded on high-level intelligence generated by a machine processable semantic information layer. Applications may be around, which exceed the claim of a search machine and might give rise to a richer vision of a semantically grounded human-machine interaction. Of particular interest appear all aspects of navigational intelligence, as they form the core of interactivity in any hypermedia application context.

This paper discusses semantic approaches to interactivity in an open hypermedia system context. The application examples originate in experiences of educational content management. Our approach starts from the primary, most often violated principle of educational content applications, the strict separation of structure, logic, content and design, as it can be achieved by applying XML-technologies in a rigorous fashion. Here it should be noted that hyperlinks, from our view, belong to structural information and therefore must not be stored within content. We will discuss a semantic representation of hyperlinks and a prototypic model for a semantic link processing along the line of this article.

This paper is organised as follows. In section 2 we introduce a translational scheme of metadata decorated content into semantic statements, including a representation of hyperlinks. Section 3 presents the concept of defining and processing linking schemes from contextual information of hyperlinks. Section 4 is dedicated to a brief introduction of the MIRaCLE, an implementation of our approach in hyper referential interactivity. Finally, section 5 gives a conclusion and an outlook on the ongoing work.

## 2    Metadata, Semantic and Hyper-Relations

### 2.1 Identifying the Semantic of Hyperlinks

The common approach of the semantic web lies in provisioning of resource descriptions and ontologies to robots such as search machines [BHL01]. Consequently, facing the current status of information in the web, a major effort concentrates on the acquisition of semantic statements describing resources and on building up ontological databases. Prior and in parallel to these recent activities the hypermedia research community has completed 30 to 50 years of research concerning the organisation and interrelation of content [OHR02], the major focus circling around the (networked) hyperlink.

An increasing number of application specific communities concerns about metadata descriptions of their information bases, acting in parallel to the semantic web initiative. Their motivations do not only derive from subject specific categorisation and retrieval, but also stem from tasks of automated content processing and context-oriented presentation.

For a vital field we concentrate in the examples given below on educational content management and related meta-data descriptors. A rich standard for the annotation of educational material has been released, the Learning Object Metadata (LOM) [LOM02].

LOM not only provides dedicated semantic descriptors of the annotated learning objects, but also includes the option of defining inter-object relations. The LOM encoding explicitly addresses issues of adaptive content processing within educational applications.

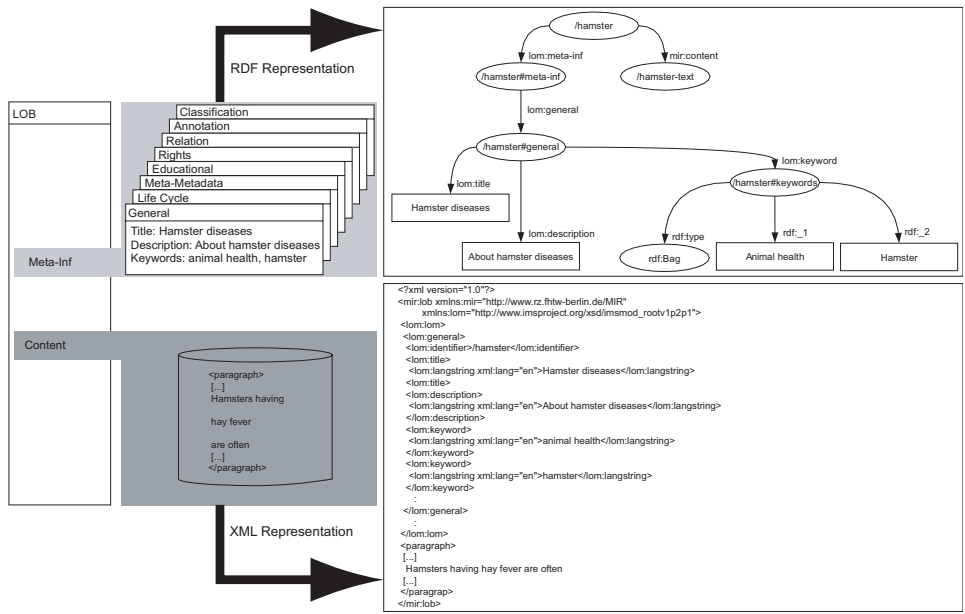

Figure 1: Representation of Metadata in XML and RDF Statements

Assuming those metadata in presence of content items a canonical semantic description is easily derived: Using RDF [LS99] representation the content object attains the role of the subject, the name of the meta descriptor forms the predicate and the value of it denotes the object. Figure 1 illustrates the different representations of the statement "this learning object is a description about hamster diseases" as in the LOM/XML schema and in RDF.

To approach a semantic analysis of hyper referential links let us recall that a hyper reference is constructed of two entities, anchors and links. Links concatenate anchors, which identify sub portions of content. In a fairly general fashion anchors can be expressed within XLink [DMO01] statements by XPointer/XPath-like expressions [DMD02, CD99], the exact formalism depending on the media type of the document. Links as well as anchors may be stored separate from document resources, e.g. in a link base.

Even though it appears rather straight forward that a semantic description of an anchor should inherit the expository statements of the underlying content, sole information inheritance remains insufficient, since a document in general may carry several, sub specific anchors. It is therefore important to provide additional specifications as can be done by the title and label tags inherent with XLink locator expressions.[1] Note that the denoted data chunks in anchors need not be of textual type. Anchors in this sense must be viewed as additional specialisations, i.e. "this resource in the context of hamster diseases carries the title of hamster having hay fever". The extraction of a semantic description of anchored resources given as a collection of inherited and dedicated statements is visualised in figure 2.

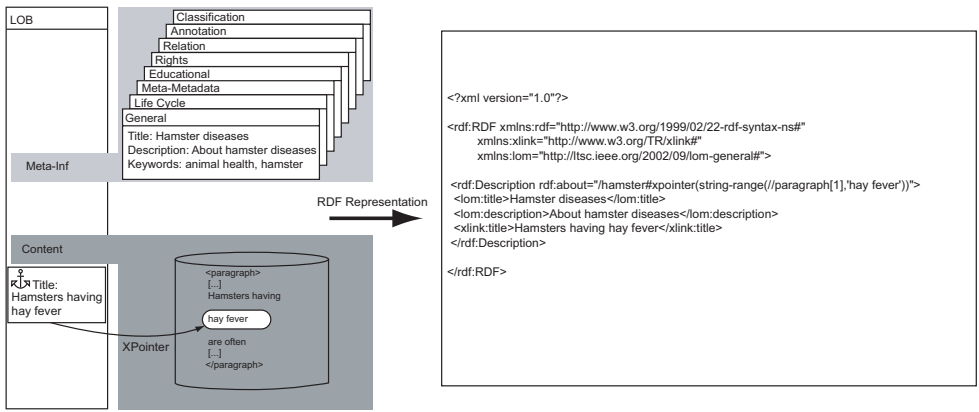

Figure 2: Gaining Additional Anchor Descriptors

Links denote relations between two or more anchors. They are directional components, uni- or bi-directional. Following the XLink arc encoding a link expression itself may carry directionless attributes, multiple titles, as well as directed descriptors, e.g. the arc attributes from, to and arcrole. A semantic of hyperlinks then naturally should build up on matching attributes, i.e. using arcroles whenever arc and direction apply, and on the linked resources. At first, this gives rise to a collection of simple statements: "This link carries the title 'For freshman'", "This link starts from the resource 'hamster having hay fever'", etc. In semantic terms statements represent linked resources. A link encodes a relation between them, which is directionally attributed by means of the arcrole. Thus at second an XLink expression gives rise to a more complex, reifying statement. A link expresses via its arcrole attribute a predicate describing the referred resources. However, in transforming this notion into a simple statement, the link resource itself remains unseen. To cure this deficit a higher order statement about statements needs to be used. Following this approach the link entity forms the subject for a statement about this relation description statement. As is visualised in figure 3 such expression reads, "Link1 denotes that resource 'Hay fever handbook' presents BackgroundInfo to resource 'Hamsters having hay fever'".

---

[1] Annotations may also appear in 'Resource' statements embedded within the content. Here a conceptual deficit within the Xlink specification becomes apparent: While using a link base with locators, attributes denoting anchored resources cannot be assigned outside actual link definitions.

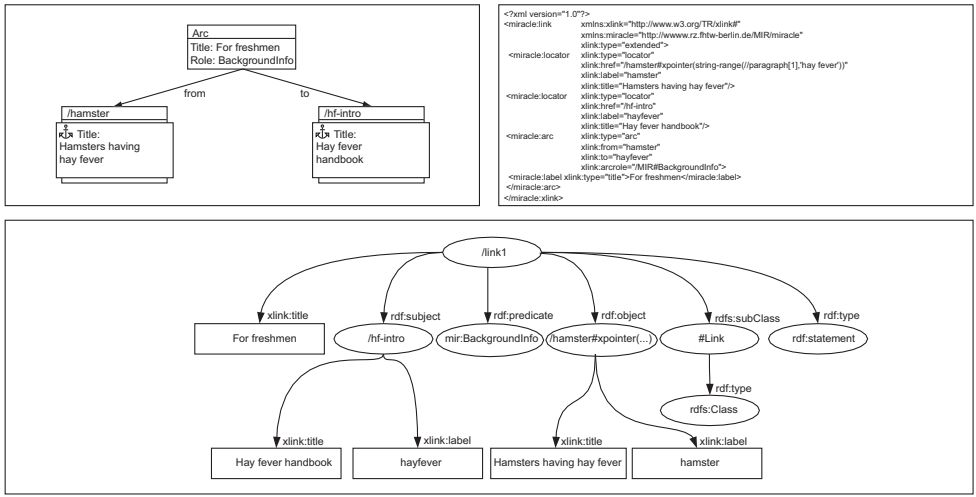

Figure 3:From XLink to an RDF Hyperlink Description

Expressing the core semantics of hyperlinks as higher order statements opens the opportunity to preserve relation to contextual information s. a. link titles etc. Viewing the approach in a rigorous semantic fashion this is indeed correct, as a link may form a resource external to content, its denoted relation being not true by itself, but an expression of contextual and personal view of the (link) author, who may be distinguished from content authors.

## 2.2 An Application Scenario

Interactivity, besides content, plays a dominant role in educational learning systems. Well-organised content can be significantly enriched or disturbed by the way links added interactivity to it. As we already mentioned in the introduction, we do not consider the definition of links as part of the content itself, but rather as part of the didactical structuring and presentation model. In particular, there should be a way to apply several linking schemes in different views to the same content.

To illustrate this argument let us return to our example: A short introductory overview on hamster diseases written in the Gaelic language is presented to a Scottish veterinarian who has significant weaknesses in Gaelic. The utmost help to him will probably be a linking, where every word is linked with the corresponding entry in a dictionary. An Irish hamster farmer with some semi-expert knowledge about his animals instead would profit most from having the medical terms linked to some background information. An Irish student of vet learning for his exam instead could appreciate all disease names being linked to some encyclopaedically knowledge for experts. And so on.

Thinking over this example we can extract the following requirements for an appropriate linking environment:

o There should be the option of applying different linking schemes to one content; thus the definition of links cannot be part of the content itself.

o Linking should adapt to the users requirements.

o A user should be enabled to adapt the linking of an application to his requirements.

o A rhetoric of linking [La89] should be present and transparent to the user.

o High-level mechanisms for defining and selecting links are needed in order to keep work of the author simple.

Some of the above requirements can be tackled with the help of Xlink/Xpath/Xpointer, but major issues remain unsolved. Introducing a semantic notion of hyperlinks raises information to a level, where authors and automated link processors within applications may be enabled to place or perform instructions effective and consistent in form and content. In section 3 we will report on our attempts in this regard.

## 2.3 Related Works

Activities of joint research in a 'semantic hypermedia' are just recently emerging. See [OHR02] for an excellent overview. Our work tackles the open problems stated herein of a semantically driven hyperlink service and the fundamental question of dealing with possible inconsistencies of knowledge, when fragments of distributed knowledge resources are hyperlinked: by encoding hyperlinks as reifying statements hyperreferential relations need not be viewed bare of context. Thus inconsistencies will arise only if contradictory statements of identical context will arrive, which then most likely cannot be ignored as artefacts.

Not coping with semantics, though, the importance of context for link processing is discussed in [El01], too. The work of the Microcosm group traditionally ranks around virtual links and anchors, which are subject to a run-time computation at the reception time. The basic idea is to dynamically add multi-destinationed links stored in a link base to (web) resources. From the hypermedia perspective these concepts are closest to our approach. In contrast to our work presented here, the group attempts to determine a link context by analyzing the anchor context, e.g. to prevent misleading links by ambiguous phrases. Their runtime environment is built on a multi agent architecture performing distributed tasks synchronised via a message passing model. This realisation completely diverges from our data-encoded approach.

The gab between the syntax oriented XLink and the semantically oriented RDF is bridged in the W3C note "Harvesting RDF statements from XLink". Here the perspective focuses on the extraction of semantic statements from XLink metadata. Since both, XLink and RDF, are capable of describing relations between resources, it is quite natural to 'harvest' semantic statements from hyperlink encodings. The arcrole attribute as the most significant metadata data about the relation of participated resources, expands in [Da00] to an RDF statement with the start resource as the subject, the end resource as the object and the value of the arcrole attribute denoting the predicate. Following our previous argumentation this approach does not formulate statements about links, but sole statements on anchors. Problems and insufficiencies of this document centered view we discussed above.

# 3 Semantic Link Context

The concept of 'not to embed', i.e. of decorating content with anchors and collecting links in a distinguished link base, gives rise to great flexibility while attaching hyperreferences to hypermedia documents. It is thereby possible to apply different linking schemes to the same content and present linking selections adaptable to the users needs. Any linking scheme, though, should follow certain demands on consistency and coherence, a rhetoric pattern for example as stated in the fundamental work of Landow [La89].

Connecting distributed knowledge resources is more than simply adding a link to a document. Having derived a semantic notion of annotated content, anchors and hyperlinks in section 2 we are now able to define a high-level scheme for collecting and processing links as members of a link-base according to application specific requirements.

 In hypermedia processing the context is an important concept [En03]. There are different contexts to recognise: The context a document appears or is to be presented in, the context of document fragments, given by its surrounding document data, and the context of a hyper reference. The latter decomposes into the source and destination context of a link, which more or less coincides with the context of the anchor fragments [HBR94], and the context of the linking entity as discussed in section 2.

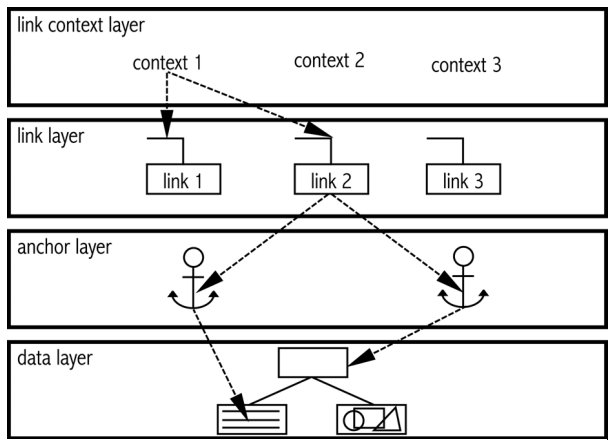

Figure 4: The MIRaCLE link layer model

In the present paper we are concerned with this semantically relevant link context. Since explicitly given by the author the context of a link need not to be determined at run time by analysing the current documents context. Link contexts are capable of articulating certain orthogonal information such as the author, the view or the proposed application of a hyper relation. To exploit these additional encodings a high-level semantic selection layer is needed to perform operations on link selections and collections based on the link context. Providing such mechanisms will enable users to steer hyperlink appearance by semantic criteria and thus interact more precise and purposeful with a hypermedia application. There are many imaginable operations like extracting links depending on their semantic role, attributes or on the relationship with their anchors or adapt them to users within personalised hypermedia applications.

The concept of a link context layer introduces a new abstraction on link collections. Within this layer we settle descriptions about a selection scheme for links following predefined semantic rules, operating on an abstract data model provided by the link layer. Link contexts neither create new links, nor new anchors. They are only responsible for the extraction of existing links stored in a link base. Those links are characterized by their descriptive properties as shown in the previous section, which combine in a selection scheme to represent a certain semantic context.

Link contexts are the upper tier in a four layered model consisting of a data, an anchor, a linking and the link context layer (s. figure 4). MIRaCLE (MIR adaptive context linking environment) is both, a formal model and a practical implementation based on the MIR system [MIR03], a runtime and storage layer presented in the following section. Each tier encapsulates certain data entities and access logics. All communication is only done between two neighbouring layers. The data layer stores raw data like text or images. Anchors are part of the anchor layer, addressing and marking chunks of data from the subordinate layer. Links connecting two or more anchors are provided by the link layer. Both, links and anchors are stored as discrete entities in a link base, offering addressability and meta data as part of an interface for the link context layer.

As shown before all semantically relevant notions from the link or anchoring layer are expressible in a formal RDF model. The link context itself is working on the model of the link layer representation and enables the user to select groups of links expressing a semantic relation.

Retrieving links means picking sub-graphs from the model. The extraction of sub-graphs could be done by an appropriate query language, like RDQL [Se02, MSR02]. Depending on the query language features one could take advantage of extended functionality like inference.

The result of such a query are statements which have the chosen links as subject and at least one predicate object pair formed by the involved anchors and the relationship existing between. Identifying the subject of the return statements as being a link gives all necessary information for further processing. An application could extract the participating anchors, verify them for being a start resource regarding the current document and visualize them, if wanted.

Syntactically the link context consists of a descriptive part and a second one containing the selection statements. The descriptive part should give at least some information about the author and purpose of the current context. We suggest using the vocabulary of Dublin Core[JP03] for the description part.

Coming back to our example on the vet student reading the text about hamster diseases let us imagine a context, which selects links providing some background information on the current topic like for example "Hamsters having hay fever". One possible link context definition is given by figure 5.

```
<?xml version="1.0"?>
<rdf:RDF xmlns:rdf="http://www.w3.org/1999/02/22-rdf-syntax-ns#"
xmlns:mir="http://www.rz.fhtw-berlin.de/MIR"
xmlns:dc="http://purl.org/dc/elements/1.1/">
<rdf:Description rdf:about="link-context1">
<dc:Creator>Mr. X</dc:Creator>
<dc:Title xmL:lang="en">Background Information</dc:Title>
<dc:Description xml:lang="en">Some continuative information
on.</dc:Description>
<mir:link-context>
<![CDATA[
SELECT * WHERE (?link, <rdf:predicate>, <mir:BackgroundInfo>) USING
rdf FOR <http://www.w3.org/1999/02/22-rdf-syntax-ns#>,
mir FOR <http://www.rz.fhtw-berlin.de/MIR#>
]]>
</mir:link-context>
</rdf:Description>
</rdf:RDF>
```

Figure 5: An example of a link context definition

The query will return all matching nodes in the graph which are the subjects of the associated RDF statements. The subjects will contain the name of the appropriate link and a statement about the connected anchors.

In the terms of our example it will return the link which expresses: "Link1 denotes that resource 'Hay fever handbook' presents BackgroundInfo to resource 'Hamsters having hay fever'". This higher order statement contains a simple statement embedding the target anchor being the subject, the predicate expressing the relation and the source anchor as the object. There are all necessary information for rendering the link into the document.

# 4 Implementing MIRaCLE on MIR

The field of educational hypermedia systems, though quite old, continues to show very active research and development activities. Numerous concepts, technologies and platforms presently are under work or design, the most prominent technological framework being XML-related. Our practical implementations rank around XML formats and technologies, as well, and rely on the more general storage and runtime platform Multimedia Information Repository (MIR) [MIR03]. Grounded on a powerful media object model MIR was designed as a universal fundament for ease in modelling and implementing complex multimedia applications.

Built on a three-tiered architecture MIR provides general support of media data handling, authentication, user and connection handling. Its core is formed by a media object database, implementing a duality of object oriented information model and relational structure. The system offers a free layer for application specific modelling of information and structures, the latter being twofold as passive structures and 'active' references, where traversal is accompanied by underlying code execution. A generic web authoring allows for immediate editing of the modelled information and structures. MIR is built as an open hypermedia system and currently supports the standards XML, Corba and JNDI. For further reading we refer the reader to [Fe01] and [FS01].

The MIR adaptive Context Linking Environment (MIRaCLE) was designed to meet all the requirements stated in section 2.2. MIRaCLE is an adaptive scheme for dynamic link decoration and generation, especially suited for internet-based teaching. The model aims to enable teachers and students define linking behaviour in semantic terms. MIRaCLE has been implemented within the MIR platform consisting of the following components:

o **Generic Anchors** are built from interfaces at all fine-grained information components. These anchor components are addressable without authoring effort.

o **Media-specific Anchors** are built upon the active reference features of the MIR system. Depending on the media type fragments of the data can be addressed with the help of mime-specific selector functions. For textual XML data the selector function is an XPointer/XPath implementation.

o **Links** are information entities, which provide references (to anchors) and metadata as needed to feed XLink expressions. Note that the MIR system automatically provides metadata exceeding the meager XLink standard (author, date, application context, …) and could provide an additional multitude without effort.

o **Link-base** resides within the MIR data store, which can be organised like a file system or grouped in application specific path spaces. The link-base offers an appropriate retrieval logic. Note that the MIR system automatically allows for structural traversal in downward and upward directions.

- o **Link Context Layer** holds and processes the semantic selection instructions (RDQL queries) stored as link contexts within the MIR database. It uses the RDF model generated by the link layer as input and returns all RDF statements matching the given RDQL query. As the result of this "semantic filtering" the appropriate subset of links is forwarded to the page builder and page renderer.

All entities, anchors, links and the activated link contexts are processed on the fly as content gets observed through a standard Web browser.

# 5 Conclusions and Outlook

In this paper we presented a semantic notion for the hyperreferential components of an open hypermedia system. In particular we reinterpreted hyperlinks as reifying statements and introduced the concept of link contexts as a semantic configuration layer for automated link processing. With the framework and implementation of the adaptive linking environment MIRaCLE we illustrated use and potential enhancements of the theoretical concept.

In future work we intend to bring those ideas and implementations into the lively context of a learning application system, as has been already started with the Hypermedia Learning Object System (Hylos) [MIR03]. More intricate semantic treatment opens up by including ontologies. We will also have an open eye on this.

**Acknowledgments**

The authors wish to warmly thank Andreas Kárpáti and Torsten Rack for many illuminating discussions. This work was supported in parts by the German Bundesministerium für Bildung und Forschung within the project Musical.